 \documentclass[final,5p,times,twocolumn]{elsarticle}

\usepackage{soul}
\usepackage{url} 

\usepackage{amssymb}
\usepackage{amsmath}
\usepackage{lipsum}
\usepackage{subcaption}
\usepackage{xcolor}
\usepackage[normalem]{ulem}
\usepackage{amssymb,amsmath,amstext,amsthm,amsfonts,amsfonts}
\usepackage[T1]{fontenc}

\usepackage{hyperref}

\journal{JINST}

\begin{document}

\begin{frontmatter}



\title{Beam test results of a fully 3D-printed plastic scintillator particle detector prototype}

\author[1]{Botao Li\corref{cor1}}
\ead{libota@student.ethz.ch}
\author[1]{Tim Weber}
\author[1]{Umut Kose}
\author[1]{Matthew Franks}
\author[1]{Johannes Wüthrich}
\author[1]{Xingyu Zhao}
\author[1]{Davide Sgalaberna}

\author[2]{Andrey Boyarintsev}
\author[2]{Tetiana Sibilieva}

\author[3,4,5]{Siddartha Berns}
\author[3,4,5]{Eric Boillat}

\author[6]{Albert De Roeck}
\author[1]{Till Dieminger}

\author[2]{Boris Grynyov}

\author[3,4,5]{Sylvain Hugon}

\author[1]{Carsten Jaeschke}

\author[1]{Andr\'e Rubbia}
  
\affiliation[1]{organization={ETH Zurich, Institute for Particle Physics and Astrophysics}, 
                postcode={CH-8093},
                city = {Zurich}, 
                country = {Switzerland}}
\affiliation[2]{organization={Institute for Scintillation Materials NAS of Ukraine (ISMA), National Academy of Science of Ukraine (NAS)}, 
                addressline={Nauki ave. 60}, 
                postcode = {61072}, 
                city = {Kharkiv},
                country = {Ukraine}}
\affiliation[3]{organization={Haute Ecole Sp\'ecialis\'{e}e de Suisse Occidentale (HES-SO)}, 
                addressline = {Route de Moutier 14},
                postcode = {CH-2800},
                city = {Del\'emont}, 
                country = {Switzerland}}
\affiliation[4]{organization={Haute Ecole d'Ing\'enierie du canton de Vaud (HEIG-VD)}, 
                addressline={Route de Cheseaux 1},
                postcode={CH-1401},
                city={Yverdon-les-Bains},
                country={Switzerland}}
\affiliation[5]{organization={COMATEC-AddiPole, Technopole de Sainte-Croix}, 
                addressline={Rue du Progr\`es 31},
                postcode={CH-1450},
                city={Sainte-Croix},
                country={Switzerland}}
\affiliation[6]{organization={Experimental Physics department, European Organization for Nuclear Research (CERN)}, 
                addressline={Esplanade des Particules 1}, 
                postcode={1211 Geneva 23}, 
                country={Switzerland}}

\cortext[cor1]{Corresponding author}

\begin{abstract}
Plastic scintillators are widely used for the detection of elementary particles, and 3D reconstruction of particle tracks is achieved by segmenting the detector into 3D granular structures. In this study, we present a novel prototype fabricated by additive manufacturing, consisting of a 5 $\times$ 5 $\times$ 5 array of 1 cm$^3$ plastic scintillator cubes, each optically isolated. This innovative approach eliminates the need to construct complex monolithic geometries in a single operation and gets rid of the traditional time-consuming manufacturing and assembling processes. 
The prototype underwent performance characterization during a beam test at CERN's Proton-Synchrotron facility. Light yield, optical crosstalk, and light response uniformity, were evaluated. The prototype demonstrated a consistent light yield (LY) of approximately 27 photoelectrons (p.e.) per channel, similar to traditional cast scintillator detectors. Crosstalk between adjacent cubes averaged 4-5\%, while LY uniformity within individual cubes exhibited about 7\% variation, indicating stability and reproducibility. These results underscore the potential of the novel additive manufacturing technique, for efficient and reliable production of high-granularity scintillator detectors.
\end{abstract}




\end{frontmatter}




\section{Introduction}
\label{introduction}

Since its invention in the early 1950s \cite{PhysRev.80.474}, plastic scintillator (PS) detectors have undergone significant advancements and are widely used in the detection of elementary particles in high-energy physics, taking advantage of its extremely fast response with time resolution in the sub-nanosecond range~\cite{Amaudruz:2012esa,Aliaga:2013uqz,MINOS:2008hdf,Joram:2015ymp,TheATLASCollaboration_2008}. By manufacturing PS detection units of different geometries, detectors are built targeting different study subjects, including time-of-flight detectors typically made of long scintillating bars \cite{Betancourt2017,Korzenev:2021mny}, neutrino active targets with tonne-to-kilotonne scale mass \cite{Amaudruz:2012esa,Aliaga:2013uqz,MINOS:2008hdf}, sampling calorimeters consisting of layers of segmented PS alternated with heavier materials such as iron and lead \cite{Allan:2013ofa}, and scintillating optical fibers \cite{Joram:2015ymp} of diameter down to 250 $\mu$m. 

In recent years, novel three-dimensional (3D) granular scintillating detectors have been developed, aiming to image charged particle tracks, electromagnetic and hadronic showers~\cite{calice}, as well as neutrino interactions~\cite{SoLid:2017ema,Sgalaberna:2017khy, Mineev:2018ekk,sfgd-testbeam-cern} by 3D tracking traversing charged particles.
For instance, a neutrino detector prototype made of approximately two million PS cubes, each of 1~cm$^3$ size and read out by three orthogonal wavelength-shifting (WLS) fibers, has been built and installed at the T2K neutrino experiment in Japan and started to collect data in 2024~\cite{ND280upgrade-tdr,nd280upgrade-press-release}. Each cube was manufactured with an injection molding technique and optically isolated from each other by a white reflective surface created using a chemical etching process \cite{Fedotov:2021ylh}.
Given the fine granularity of the detector, the trajectories of the minimum ionizing particles (MIP), such as muons, can be reconstructed with a spatial resolution ranging between 2 and 4 mm, depending on the implemented reconstruction algorithms, demonstrated by simulation studies carried out independently on a detector of similar 3D granularity structure \cite{Alonso-Monsalve:2022zlm}.

On the other hand, the requirement of a multi-tonne active volume combined with high granularity leads to complexities in the manufacturing, and significant amount of effort demanded by the assembly process \cite{Fedotov:2021ylh}.

In order to satisfy the demand of a fast production and easy assembly, Fused Deposition Modeling can be a promising candidate for 3D-printing PS-based scintillator\cite{fdm}. A first proof of concept has been reported in \cite{Berns:2020ehg} in 2020 with a 1 cm$^3$ single transparent PS cube. The method has been further developed and validated in 2022 \cite{3DET:2022dkw}, where a 3 $\times$ 3 voxel layer has been 3D-printed together with the optical isolation walls between active voxels. MPPCs were coupled directly to the scintillator volume for the optical measurement. A satisfying light yield (LY) and low cube-to-cube crosstalk was measured. Furthermore, a 3D-printed PS bar of 5 cm length has been manufactured in order to characterize the optical attenuation of the 3D-printed scintillator blocks. A 19 cm technical attenuation length was measured, which is sufficient for fine-granularity scintillator detectors \cite{3DET:2022dkw}. However, some air gaps and bubbles were visible in this prototype and the geometrical precision was not optimal.

More recently, a novel manufacturing method named Fused Injection Modeling (FIM) has been developed based on the traditional FDM techniques~\cite{FIM}, enabling automatic 3D-printing of a single, large block of fine-granularity, optically-isolated PS structure, possessing good geometrical tolerance, high transparency, as well as precise fabrication of the hollow structure for the purpose of readout fiber placement. FIM adopts the capability of manufacturing complex geometries of FDM, by 3D-printing an optically-reflective frame of the desired voxel shape and size, and the fast production speed and high part density of injection molding, by rapidly pouring melted PS into the empty cavity that accurately shapes the scintillator geometry \cite{FIM}.

\section{Manufactured particle detector prototype}
A first prototype, the "SuperCube", was manufactured with FIM in 2023 and consists of 5 $\times$ 5 $\times$ 5 optically-isolated PS voxels, each of 1 cm$^3$ size. The detailed manufacturing process can be found in~\cite{FIM}. Each voxel of the SuperCube is read out by two orthogonal WLS fibers enabling a 3D-reconstruction of the trajectory of the traversing charged particle. 
The 2D-readout scheme was adopted only as a temporary technique compromise for this specific small prototype. An automatic fabrication of the hole for a third projection channel perpendicular to the current two views is under development, which will improve the multi-track reconstruction by reducing the amount of the artificial fake hits at the reconstruction stage~\cite{ND280upgrade-tdr}.
The holes for the fiber insertion were formed during the filling process by polished metallic rods easily extracted once the scintillator had solidified. The prototype was first tested using cosmic muons in 2023, with results reported in~\cite{FIM}. For performance benchmarking, the prototype was evaluated together with another prototype of identical voxel dimensions, 
manufactured using the standard cast polymerization method \cite{glued_cube}. Both prototypes demonstrated a comparable LY of about 28 p.e./channel 
when read out using Hamamatsu MPPC 13360-1325CS with $\sim$25\% nominal photon detection efficiency (PDE) \cite{hamamatsu:mppc}. However, 
the additive manufacutured prototype showed higher cube-to-cube light crosstalk (($\sim$3.8\%)), attributed to the differences in the reflective materials. Despite of this, the optical isolation of each cube is comparable with that of the state-of-art instrumented 3D-granularity neutrino detector~\cite{sfgd-testbeam-cern}, where a typical crosstalk rate of $\sim$3\% was measured. This level of crosstalk is acceptable for the physics purpose. The feasibility of manufacturing better reflective material with FIM technique is currently under development.

To further validate its performance, the same prototype was exposed to a beam of charged particles at CERN in August 2023. The experimental setup and the detailed results, including measurement of LY,  cube-to-cube optical crosstalk, and spatial LY non-uniformity within a single PS cube, are discussed in the following sections.

\section{Experimental Setup}
\begin{figure}
	\centering 
	\includegraphics[width=0.4\textwidth]{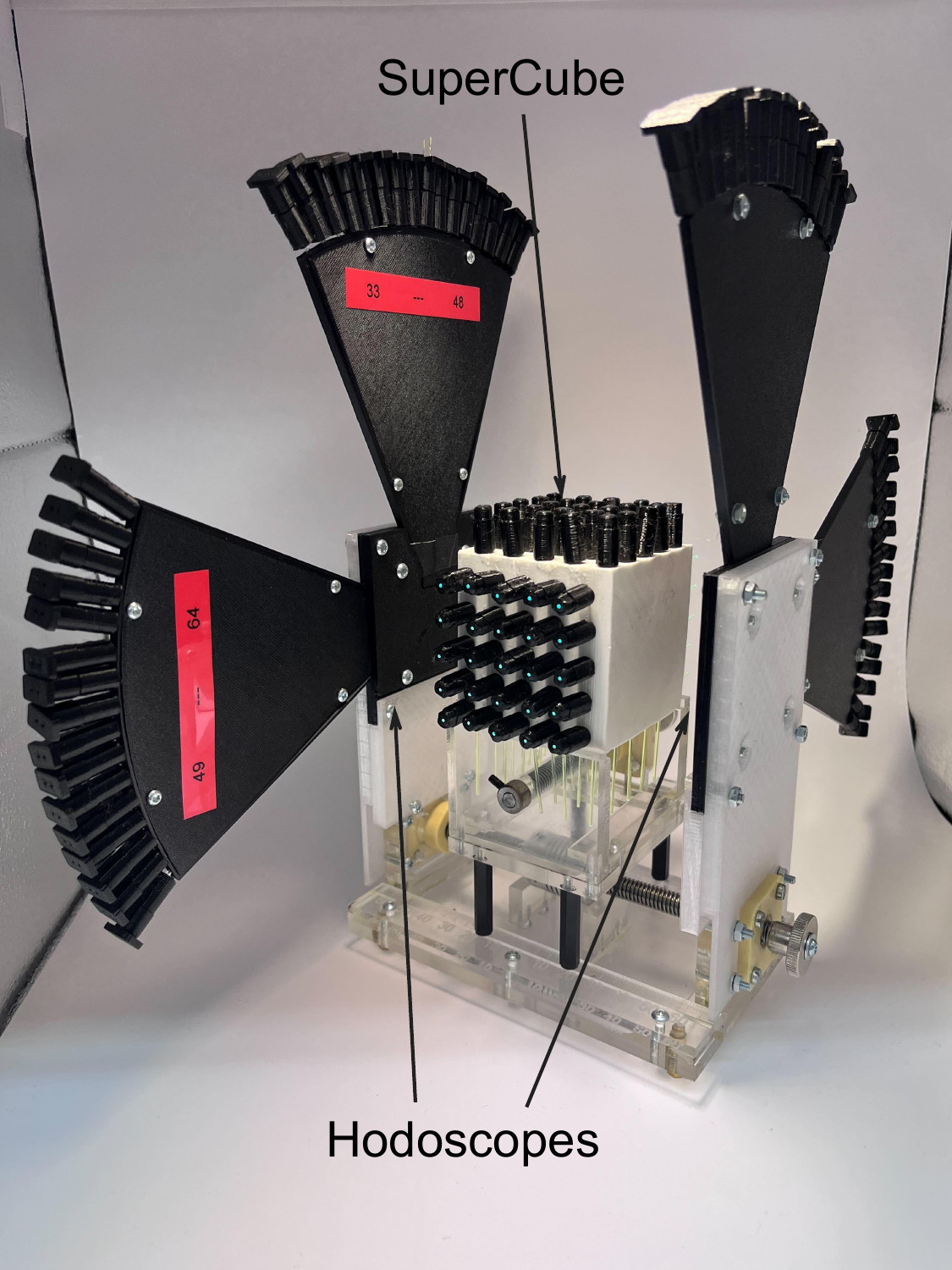}
	\caption{The prototype setup. The SuberCube was placed at the center, with WLS fibers inserted. The hodoscopes were installed at both sides of the SuperCube providing finer resolution for traversing tracks. }
	\label{fig:setup}
 
\end{figure}

The detector setup, shown in Fig.~\ref{fig:setup}, consists of the SuperCube, complemented by two scintillating fiber hodoscopes for high-resolution tracking. The PS prototype was fixed at the middle of the frame, with two hodoscopes symmetrically positioned 6 cm from the cube center on either side. Each hodoscope was composed of two orthogonal layers of scintillating fibers. This arrangement ensures sub-millimeter spatial resolution in two dimensions on the plane perpendicular to the test beam.


\subsection{Fully 3D-printed plastic scintillator particle detector}
\label{sec:supercube}
The scintillating cubes of the SuperCube were read out 
using 50 Kuraray Y11 double cladding WLS fibers~\cite{kuraray-catalogue-wls} of 1 mm diameter, and 9 cm length.
These fibers were inserted through the orthogonal holes, which were offset from the cube center with 2.5 and 3 mm distance from two sides of the cube walls respectively~\cite{FIM}. The fibers collected the scintillation light generated by traversing charged particles, enabling 3D reconstruction of particle tracks.
Both ends of each fiber were polished and one end coupled 
to a Hamamatsu MPPC 13360-1325CS\cite{hamamatsu:mppc} of 1.3 $\times$ 1.3 mm$^2$ active area, by 3D printed black connectors (male and female). The coupling was enhanced by soft elastic foams, pushing the MPPC against the fiber end. The MPPCs used in this setup had a nominal photon detection efficiency (PDE) of 
25\% and the pixel pitch of 25 $\mu$m.




\subsection{Hodoscopes}
\begin{figure}
	\centering 
	\includegraphics[width=0.3\textwidth, angle=0]{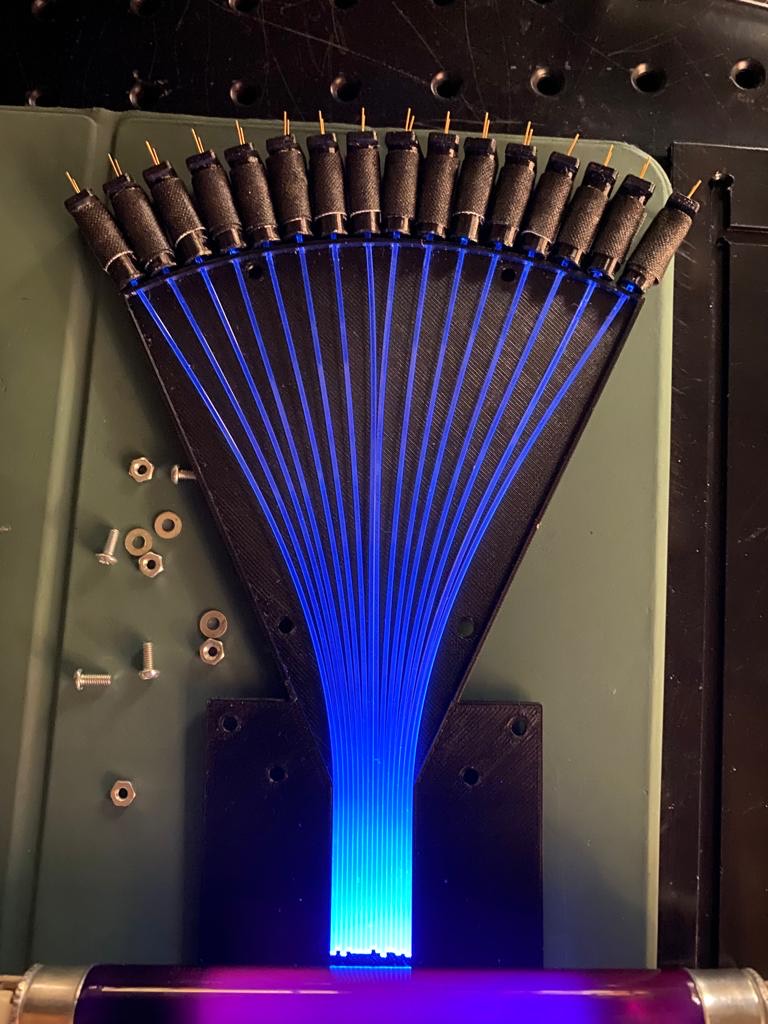}	
	\caption{One hodoscope piece under UV lamp with top cover removed} 
	\label{fig:hodoscope}
\end{figure}

Two auxiliary hodoscopes were assembled to provide sub-mm resolution for the charged particles passing through the center area of the SuperCube. This allows to study the LY non-uniformity as a function of particle position within a single scintillator cube. Each hodoscope consists of 32 1 mm square single-cladding Kuraray scintillating fibers of 17 cm length. The scintillating fibers were split into two groups of 16 fibers aligned orthogonally to each other and perpendicular to the beam direction, covering a 1.6 $\times$ 1.6 cm$^2$ area. Such configuration provides sub-mm resolution to the entering and exiting point of the particle passing through the SuperCube. Fig.~\ref{fig:hodoscope} shows one component of the hodoscope with top black cover removed and illuminated by a UV lamp. 
All the scintillating fibers were polished at both ends and read out by MPPCs, the same type as described in Sec.~\ref{sec:supercube}. A typical LY of about 14 p.e./channel/MIP was measured in preparatory tests prior to the beam tests. The hodoscopes were later integrated onto the mechanical frame ensuring good alignment between the centers of the hodoscopes and the SuperCube.

\subsection{Electronics}
A total of 50 SuperCube MPPC-channels and 64 hodoscope MPPC-channels were separately read out by two 64-channel CAEN DT5202 front-end boards (FEB)~\cite{caen-FERS-board} via 2m-long micro-coaxial cables. The analog charge signals generated by the MPPCs, proportional to the number of detected photons, were digitized by the FEBs, and converted to analog-to-digital converter (ADC) units 
by measuring the charge value corresponding to the highest signal peak.
The bias voltage for MPPCs was set at approximately 57 V, as recommended by the manufacturer. The FEBs can be operated in both self-trigger or external trigger mode with a fast readout ensuring an automatic event recording and triggering coincidence. 
For each channel, build-in charge discriminators were used to suppress noise triggering.
The FEB (Board A) reading the SuperCube channels operated under the self-trigger mode. All SuperCube channels were used for event triggering to ensure sufficient statistics. The other board (Board B) was connected to board A via an ethernet switch and started the data acquiring process of all the hodoscope MPPC channels when receiving an external trigger from the board A. 
This trigger scheme resulted in two types of recorded tracks: tracks passing through only the SuperCube and tracks producing signals in both the SuperCube and hodoscopes.
The first set of events were used to study the LY and the crosstalk of the SuperCube with high statistics, while the latter one was used to study the LY uniformity as a function of the passing position of the traversing particle within a cube. The hodoscopes allowed 
precise scanning of the cube's front and rear cross-sections with a 1 mm pitch 
in both directions perpendicular to the test beam axis.

\subsection{Test beam}
The beam test was conducted at the T9 beamline of the CERN Proton-Synchrotron (PS) East Area facility 
which is capable of providing muons, electrons and charged hadrons up to 15 GeV/c.
The SuperCube was tested parasitically at the down stream location to the ENUBET prototype, a sampling calorimeter with multilayered configuration \cite{ENUBET}, during runs with 10 GeV muon and hadron beams.

The acquired data, predominantly consisted of minimum-ionizing muons, were used to study the performance characterization of the detector.
\section{Beam test results}

\subsection{Data processing}
\label{sec:data_preprocession}
During the beam test, a total of 12,190,778 triggered events were recorded. Because of the triggering scheme, these events contain either single or multiple particle tracks propagating along various directions. 
To characterize the optical performance of the SuperCube, strict selection criteria were applied to isolate events with a clean, single track topology. 
For an event to pass the single-track selection, the following conditions were required: 
in each layer of the SuperCube perpendicular to the beam, there must be a unique channel measuring a signal exceeding 500 ADC in both projections.
This threshold ($\sim$8 p.e.) was determined based on a previous measurement of the SuperCube with cosmic muons~\cite{FIM}, where the most probable value of a typical MIP signal was found to be $\sim$28 p.e.. 
This selection ensures the single track topology in the selected events.
Additionally, since the LY of each single cube is approximately proportional to the energy deposited by a charged particle traversing its active volume, a further straight-track selection was applied.
This selection identified straight particle tracks passing through all layers of the SuperCube at roughly similar positions. These tracks propagated $\sim$1 cm in each of the cube and allowed for a more precise estimation of the energy deposition within the detector. 
After applying both the single-track and straight-track selections, 1,180,787 (9.6\%) events were retained for characterizing the SuperCube performance. 
It is worth noting that the selected dataset still contain the signals from all the 50 SuperCube channels, including not only the high charge MIP signals, but also the low charge crosstalk signals and the noise/dark count signals as well.

\begin{figure}[t!]
	\centering 
	\includegraphics[width=0.45\textwidth, angle=0]{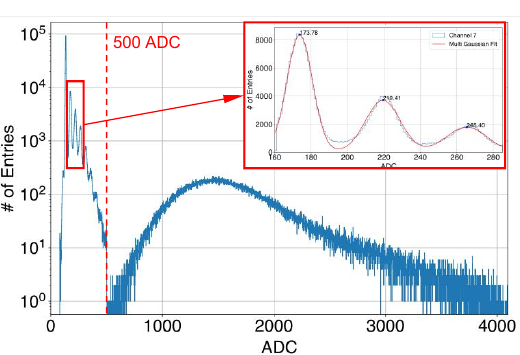}	
	\caption{The measured ADC distribution of the selected events of a randomly chosen channel 7 as an example. The low charge region (<500 ADC) shows a clear multi-peak structure, where the first highest peak shows the channel pedestal, and the following peaks correspond to each p.e. response. The zoom-in plot shows a multi-Gaussian fit of the first three p.e. peaks above the pedestal. The high charge (>500 ADC) region shows the MIP signals measured by this channel, where the channel LY was obtained.} 
	\label{fig:gainfit}
\end{figure}
\subsection{Light yield}
In order to study the LY of the SuperCube, the raw data, recorded in ADC units, were converted into photoelectrons (p.e.), which represent the primary electrons originated via photoelectric effect by the visible photons impinging on the MPPC active area. Within the dynamic range of the MPPCs, the two units are connected by a linear correspondence:
\begin{equation}
[ADC] = g * [p.e.] + p.
\end{equation}
Here $g$ is the gain of the MPPC in unit of [ADC/p.e.], and $p$ is the pedestal.
For each channel, the conversion factors, $g$ and $p$, were derived directly from the measured ADC distribution of the selected events. 
Figure~\ref{fig:gainfit} illustrates the ADC distribution of channel 7, chosen randomly as an example.
The distribution exhibits a distinct multi-peak structure below 500 ADC and a broad, single-peak distribution above the threshold.
The region below 500 ADC, defined as the low charge region, mainly consists of crosstalk signals and MPPC dark counts, collected when the cube measured by this channel was not directly hit by a charged particle, while the region above 500 ADC represents signals generated by real particle hits, and was defined as the high charge region.

The low charge (<500 ADC) region was used for the channel calibration. The first and highest peak shows the pedestal, while the subsequent peaks correspond to the responses of 1 p.e., 2 p.e., 3 p.e., and so on.
The gain of the MPPC is obtained from the distance between adjacent peaks. The inset plot of Fig.~\ref{fig:gainfit} presents the multi-Gaussian fit of the first three peaks above the pedestal of the distribution, with the peak positions obtained. Furthermore, a linear fit was applied to the peak positions, and was then used to determine 
the MPPC gain and pedestal. All the 50 channels were independently calibrated following the same scheme. The ADC-to-p.e. conversion was applied in the following analysis.

\begin{figure}[t!]
	\centering 
	\includegraphics[width=0.45\textwidth, angle=0]{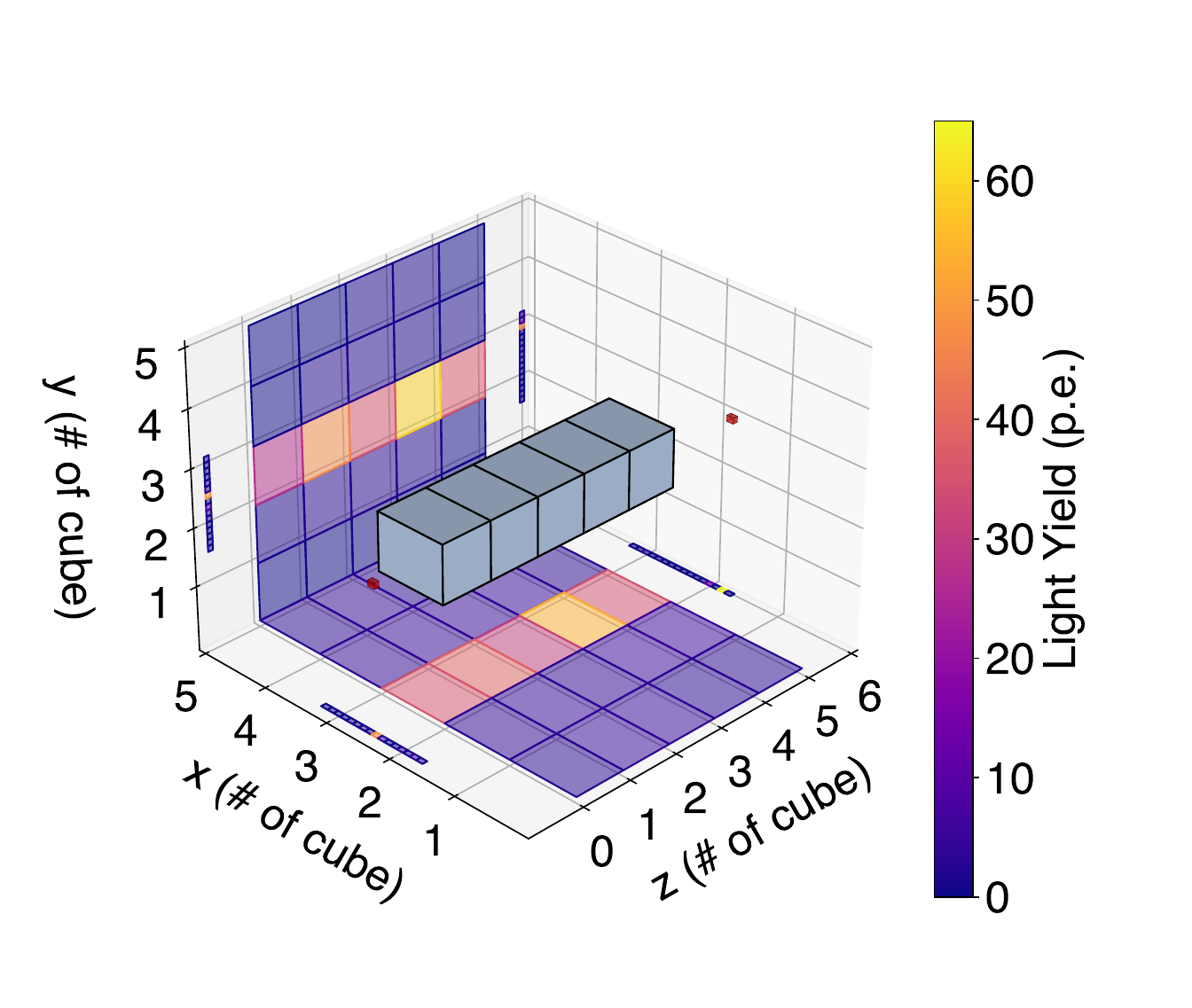}	
	\caption{Reconstructed straight track passing through both hodoscopes and the SuperCube. The fiber channel hits of both the SuperCube and the hodoscopes are shown by the 2D projections. The reconstructed SuperCube hits are represented by the large 3D cube (grey), and that of the hodoscopes represented by the small cube (red), with their size corresponding to the respective spatial granularity of two detectors. The 2D projections shows all the hits both above and below the threshold cut, while in the 3D view the crosstalk hits below the threshold were removed for a better visualization.} 
	\label{fig:event_display}
\end{figure}

For each selected event, the selection procedure decribed in Sec.~\ref{sec:data_preprocession} ensured the existence and uniqueness of a single, straight, and through-going MIP track in each of the two 2D projections of the SuperCube. Consequently, the reconstruction of a single 3D particle tracks is straightforward on an event-by-event basis.
Fig.~\ref{fig:event_display} shows the track of a particle detected both in the hodoscopes and the prototype. The hodoscope hits were reconstructed simply by matching the channels with the highest LY above the threshold in each of the 2D projections.

\begin{figure}[t]
	\centering 
	\includegraphics[width=0.4\textwidth, angle=0]{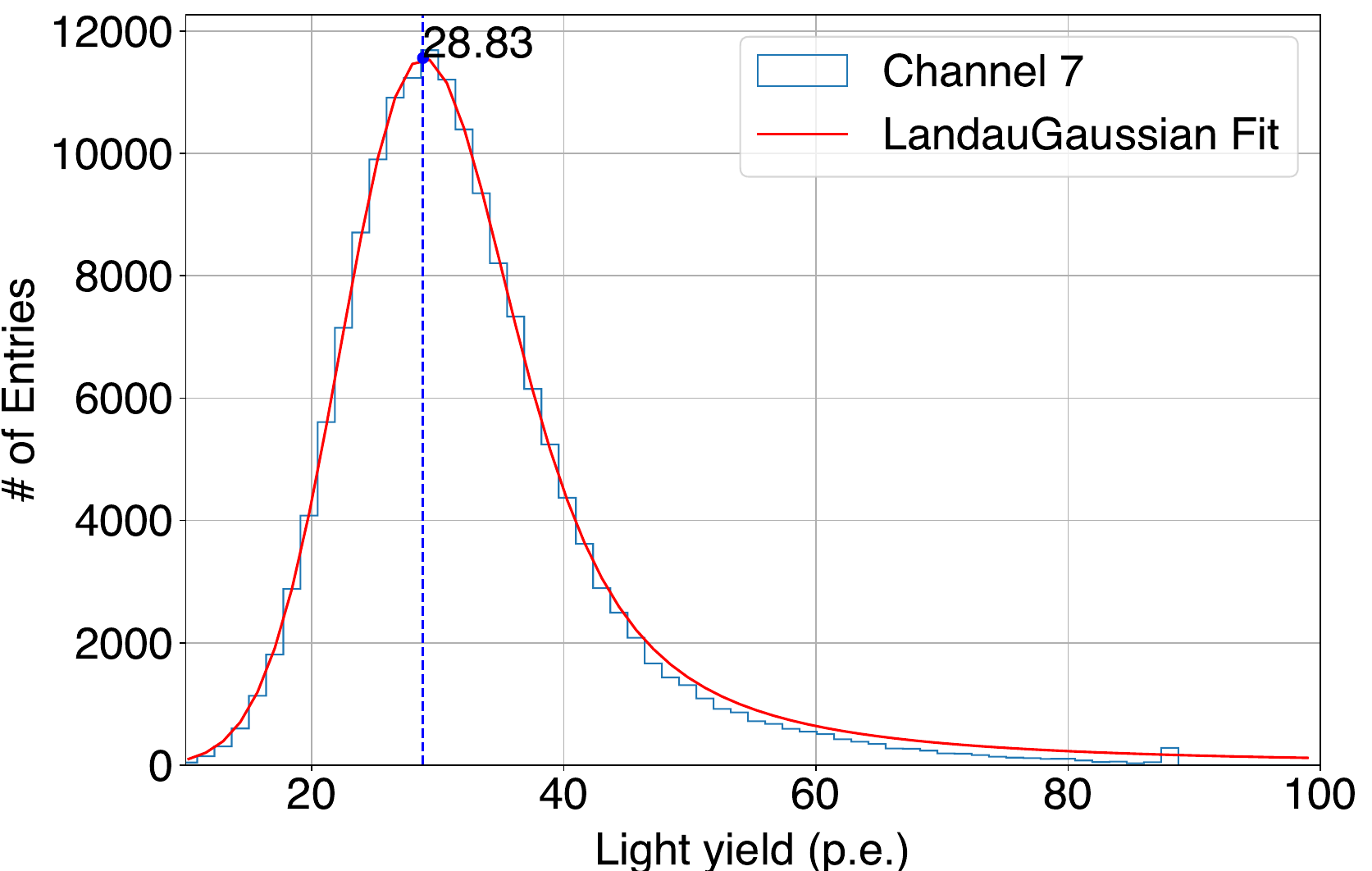}	
	\caption{Light yield distribution of one example cube channel, fitted by a Landau-Gaussian distribution.} 
	\label{fig:light_yield}
\end{figure}

The high charge region (>500 ADC) in Fig.~\ref{fig:gainfit} shows the LY distribution generated by MIPs traversing a cube measured by the channel.
The corresponding distribution after the ADC-to-p.e. conversion is illustrated in Fig.~\ref{fig:light_yield}.
The distribution was fitted with a Landau function with a Gaussian smearing to find the peak value characterizing the LY of this specific channel. The function was defined as:
$$f(x;A, \mu, \sigma_G, \sigma_L) = A\int L(u;\mu, \sigma_L)G(x-u;0,\sigma_G)du,$$
where $A$ is the overall normalization, $L$ is the Landau distribution with most probable value $\mu$ and scale parameter $\sigma_L$, and $G$ is the Gaussian smearing with zero mean and standard deviation $\sigma_G$. A typical LY of 28.8 p.e./MIP of this specific channel was obtained from the peak position of the fitted function.
Among all the 50 SuperCube channels, an average LY of 28.1 p.e./channel/MIP was measured, with a standard deviation of 3.7 p.e./channel/MIP among different channels.

\subsection{Optical crosstalk}
\begin{figure}[h]
	\centering 
	\includegraphics[width=0.5\textwidth, angle=0]{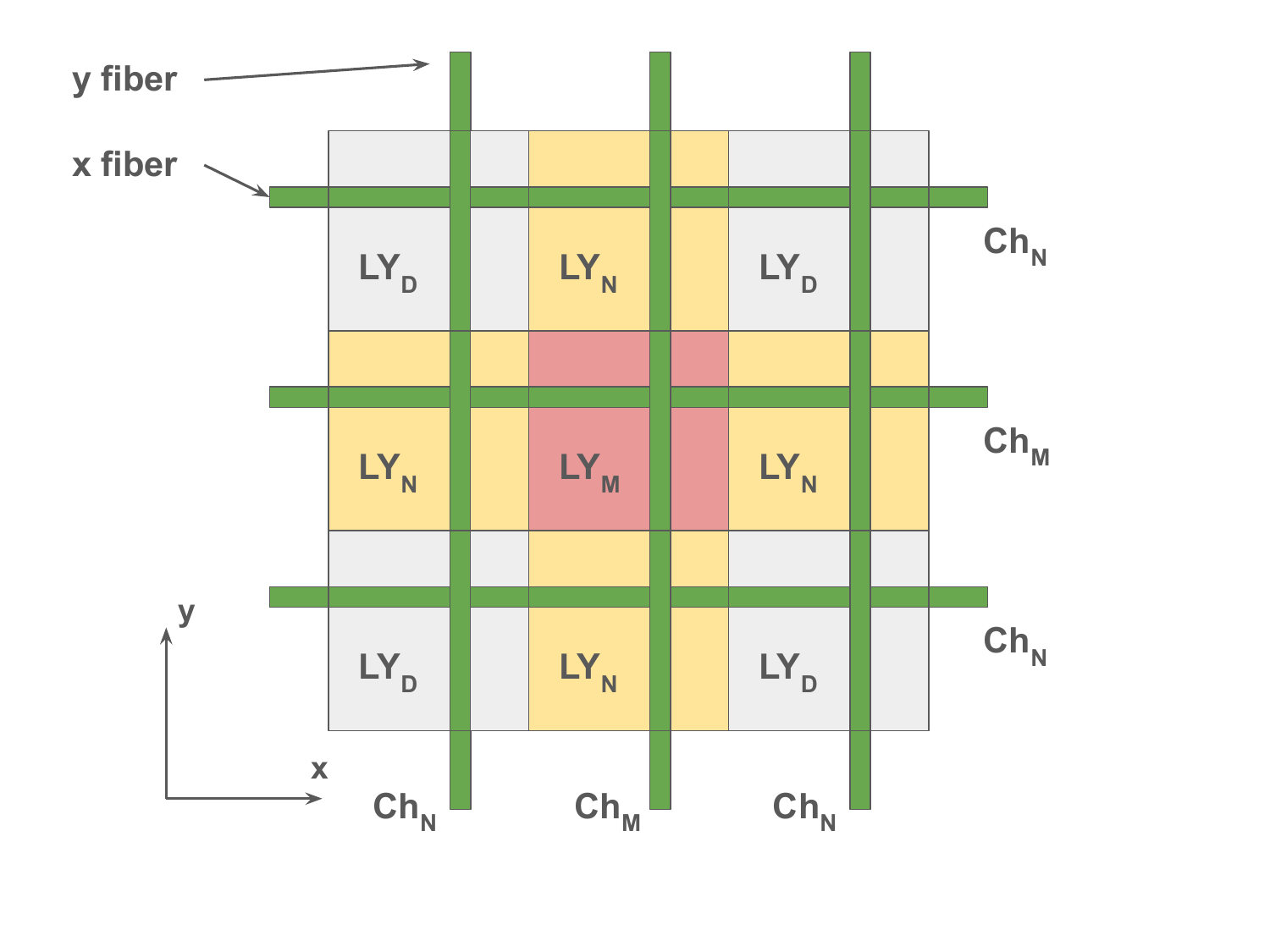}
	\caption{Sketch illustrating the main cube (red) hit by a particle propagating along z axis, and four adjacent crosstalk cubes (yellow) in x-y plane perpendicular to the test beam axis, as well as the WLS fibers along x and y axis. $LY_M$ represents the light left in the main cube after the leakage, when $LY_N$ represents the light leaked into an adjacent cube. $LY_D$ is the negligible light leakage into the diagonal cubes. The readout channel of the main cube is $Ch_M$, while the readout channel of the crosstalk cubes are $Ch_N$.} 
	\label{fig:xtalk_draft}
\end{figure}

Given the non-negligible light transmission of the white cube reflector, a fraction of the scintillation light produced in one cube can leak into adjacent cubes and be captured by their corresponding WLS fibers. This phenomenon, referred as cube-to-cube optical crosstalk, reduces the resolution of single-hit reconstruction and results in "thicker" particle track images. To quantify this effect, a dedicated method was implemented, 
as shown in Fig.~\ref{fig:xtalk_draft}.
Here, we analyzed the cube-to-cube crosstalk using 
the x-y view of a hit cube and its nine neighbors, each read out by two orthogonal WLS fibers along x and y directions. The center cube, highlighted in red, is assumed to be traversed by a MIP 
along the z-axis. 
The crosstalk is expected 
in the four neighboring cubes (colored in yellow), which share a face with the central cube. The crosstalk in the diagonal cubes is negligible. 
The LY measured from WLS fiber channels of the hit cube and neighboring cubes is represented by $Ch_{\text{M}}$ and $Ch_\text{N}$, respectively.  
The LY remaining in
the hit cube is denoted as $LY_\text{M}$, while the LY leaking into one of the adjacent cubes is represented by $LY_\text{N}$ (Equal sharing of the crosstalk through each of the six faces of a single cube is assumed). 

The cube-to-cube crosstalk through one face is expressed as $R_\text{crosstalk} = LY_\text{N}/LY_\text{M}$, and can be estimated using the LY measured from fibers of adjacent cubes. On the other hand, the fibers of the main cube also pass through two of the adjacent cubes, collecting the crosstalk at the same time. Thus, further corrections need to be considered instead of simply calculating the ratio of the fiber LY $Ch_\text{N}/Ch_\text{M}$.
Assuming the hit cube is not at any edge of the prototype, the LY measured from the main channel can be derived as:
$$Ch_\text{M} = \frac{1}{2}(LY_\text{M} + 2LY_\text{N}) + \frac{1}{2}\cdot 2LY_\text{N} = \frac{1}{2}(LY_\text{M} + 4LY_\text{N}).$$
Here, the first term 
represents the LY in the hit cube ($LY_\text{M}$) combined with the crosstalk from two adjacent cubes along the z-axis, while the second term accounts for the crosstalk collected from the neighboring cubes sharing the same fiber channel. The factor $1/2$ arises from the equal sharing assumption. Similarly, the LY of a neighbor channel $Ch_\text{N}$ is simply:
$$Ch_\text{N} = \frac{1}{2}LY_\text{N}.$$ 
The measured ratio of LY between the neighboring and main channels is defined as:
$R_\text{measure} = Ch_\text{N}/Ch_\text{M}$, using this ratio, the crosstalk can be calculated as a function of $R_\text{measure}$:
\begin{equation}
R_\text{crosstalk} = \frac{LY_\text{N}}{LY_\text{M}} = \frac{R_\text{measure}}{1-4R_\text{measure}}. \label{eq:xtalk}
\end{equation} 
When the hit cube is located at the edge of the prototype, the compensation factor 4 in Eq.~\ref{eq:xtalk} should be decreased, i.e. there is either less crosstalk from the adjacent cubes along the particle direction, or less crosstalk collected by the main channel $Ch_\text{M}$.

\begin{figure}[h!]
	\centering 
    \begin{subfigure}{0.4\textwidth}
        \includegraphics[width=\textwidth, angle=0]{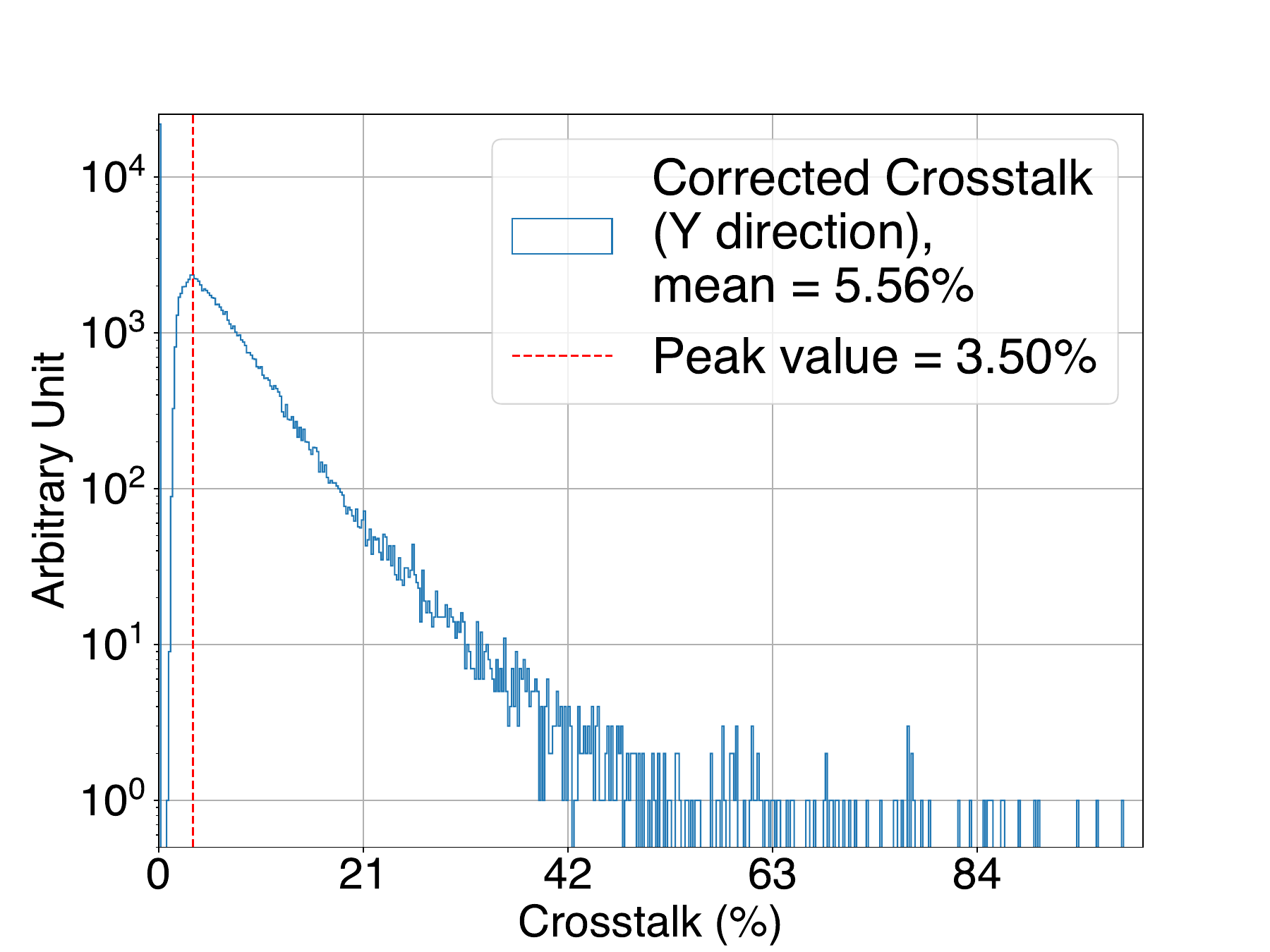}	
        \caption{y fiber channels}
    \end{subfigure}%
    \newline
    \begin{subfigure}{0.4\textwidth}
        \includegraphics[width=\textwidth, angle=0]{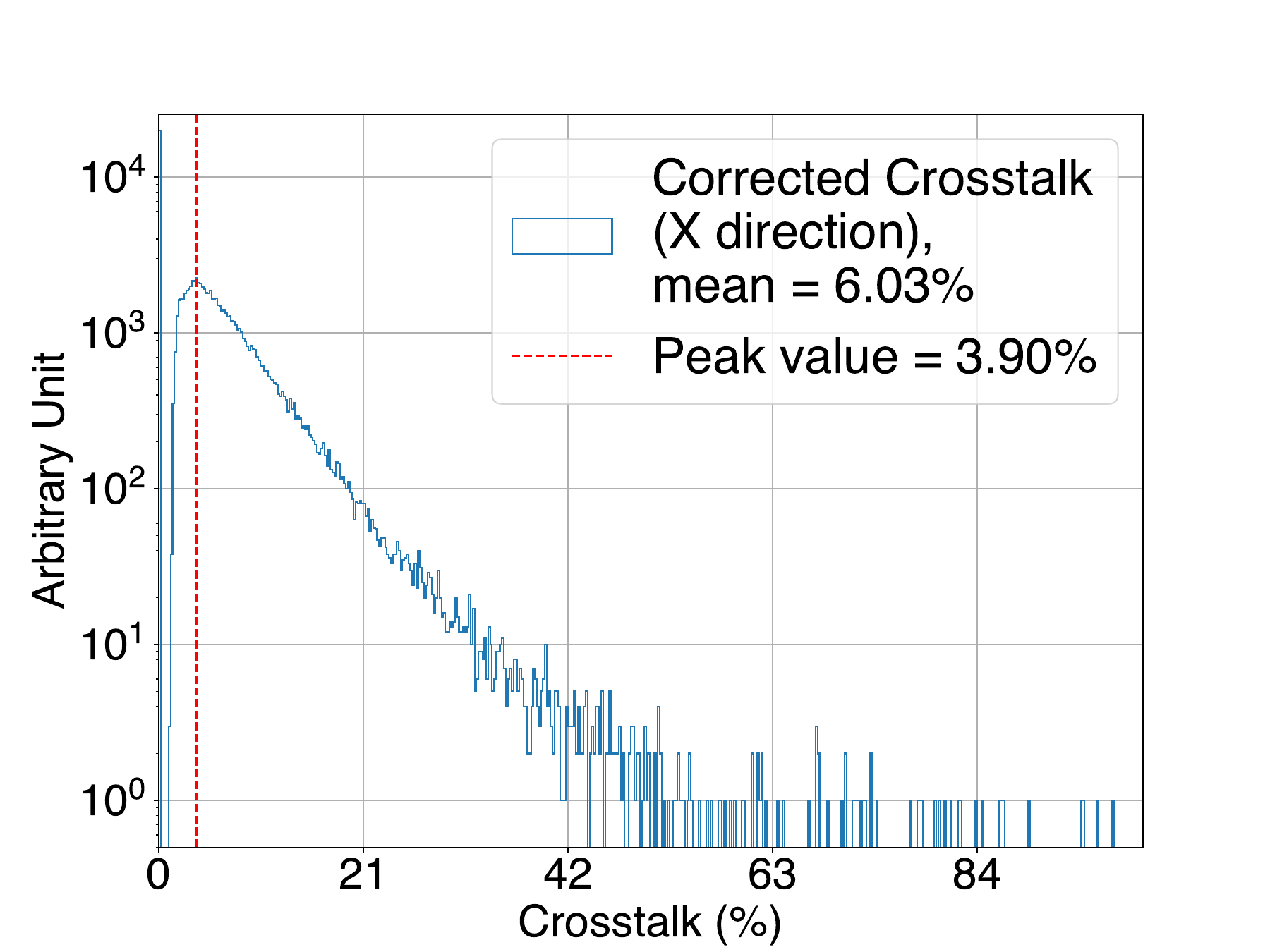}	
	    \caption{x fiber channels}
    \end{subfigure}
	\caption{The crosstalk distribution of x fiber channels and y fiber channels. } 
	\label{fig:xtalk}
\end{figure}

A high LY cut ($>$100 p.e./two fibers) was applied to the main hit cube when calculating the crosstalk, in order to ensure larger-than-pedestal readout in the neighboring channels. The analysis was restricted to the central 3 $\times$ 3 $\times$ 3 cubes region 
to avoid possible boundary effects. Figure~\ref{fig:xtalk} illustrates the distribution of the corrected cube-to-cube crosstalk, calculated using Eq.~\ref{eq:xtalk}, for the channels along x and y axis respectively. The distributions peak at approximately 3.5\% and 3.9\%, respectively. The sharp peak at 0  corresponds to events where neighboring channels recorded only pedestal signals. The long tail of the distribution with high crosstalk is likely coming from \text{$\delta$}-ray events. These results are consistent with the previous measurements with cosmic muons \cite{FIM}, and comparable to state-of-art PS detectors with similar designs \cite{sfgd-testbeam-cern}. 

\subsection{Non-Uniformity in light response}

We further investigated the LY non-uniformity as a function of the ionization position within a single cube. As described earlier, the two hodoscopes with active area of 1.6 $\times$ 1.6 cm$^2$ were placed at both sides of the prototype to cover the central row of cubes along the beam axis.
A threshold at 300 ADC (about 4 p.e.) was applied to the hodoscope events to enhance the selection purity by rejecting the low LY events, which are likely due to background signals. 
Tracks passing through all five cubes in the central row and both the hodoscopes were selected for analysis. By applying linear fits to the two hodoscope hits,  tracks were reconstructed with higher spatial resolution. The middle point of the track segment in each cube was taken as the interacting point within the x-y plane perpendicular to the beam direction. The cross section of a single cube was split into 36 (6 $\times$ 6) bins, ensuring coverage for the high angle track segments.
For each cube, the measured LY non-uniformity maps were generated separately for the two readout channels, perpendicular to the beam axis.
\begin{figure}[h!]
	\centering 
    \begin{subfigure}{0.4\textwidth}
	   \includegraphics[width=\linewidth]{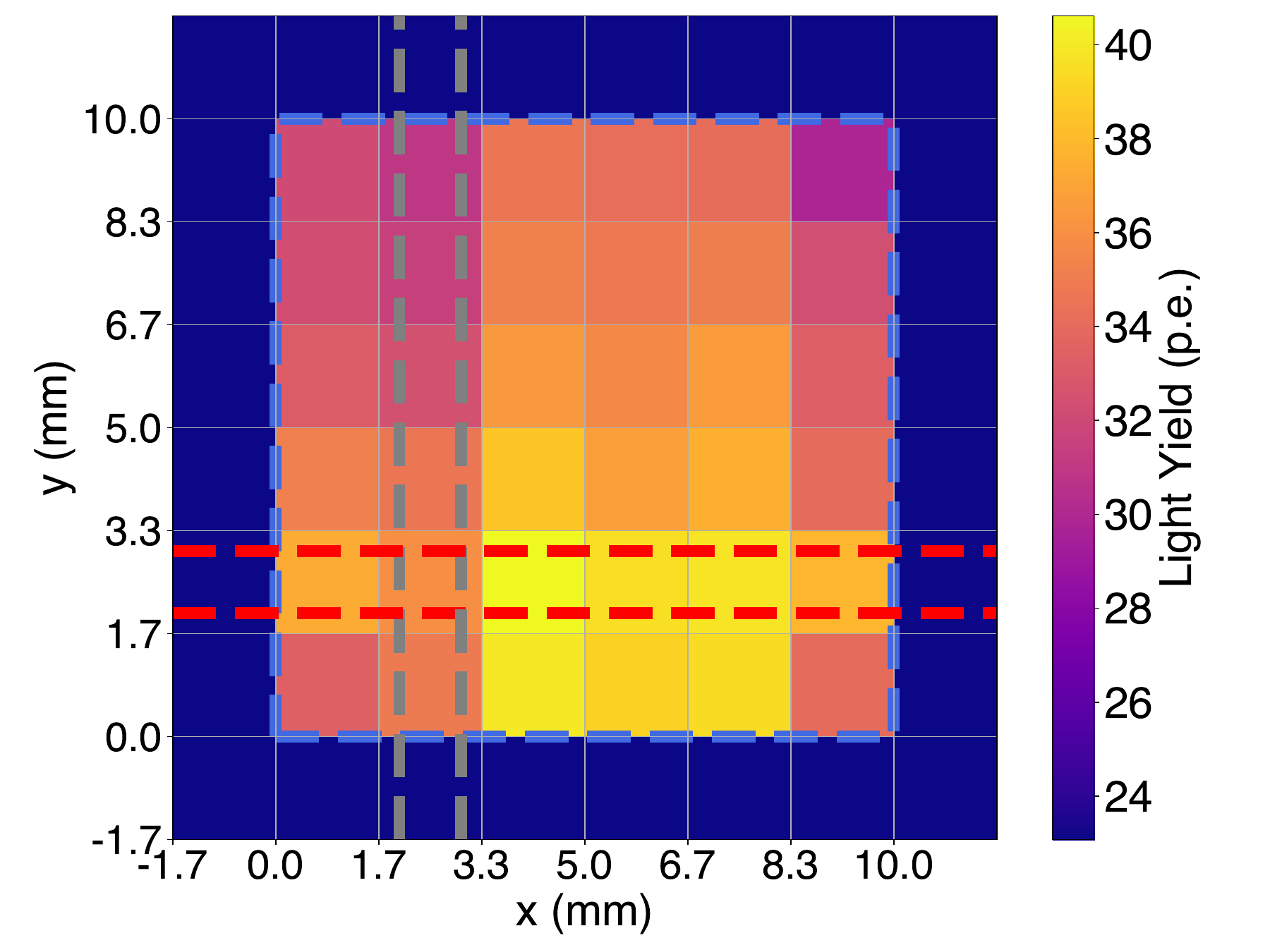}	
        \caption{The readout of the WLS fibers along the X axis.}
    \end{subfigure}%
    \newline
    \begin{subfigure}{0.4\textwidth}
        \includegraphics[width=\linewidth]{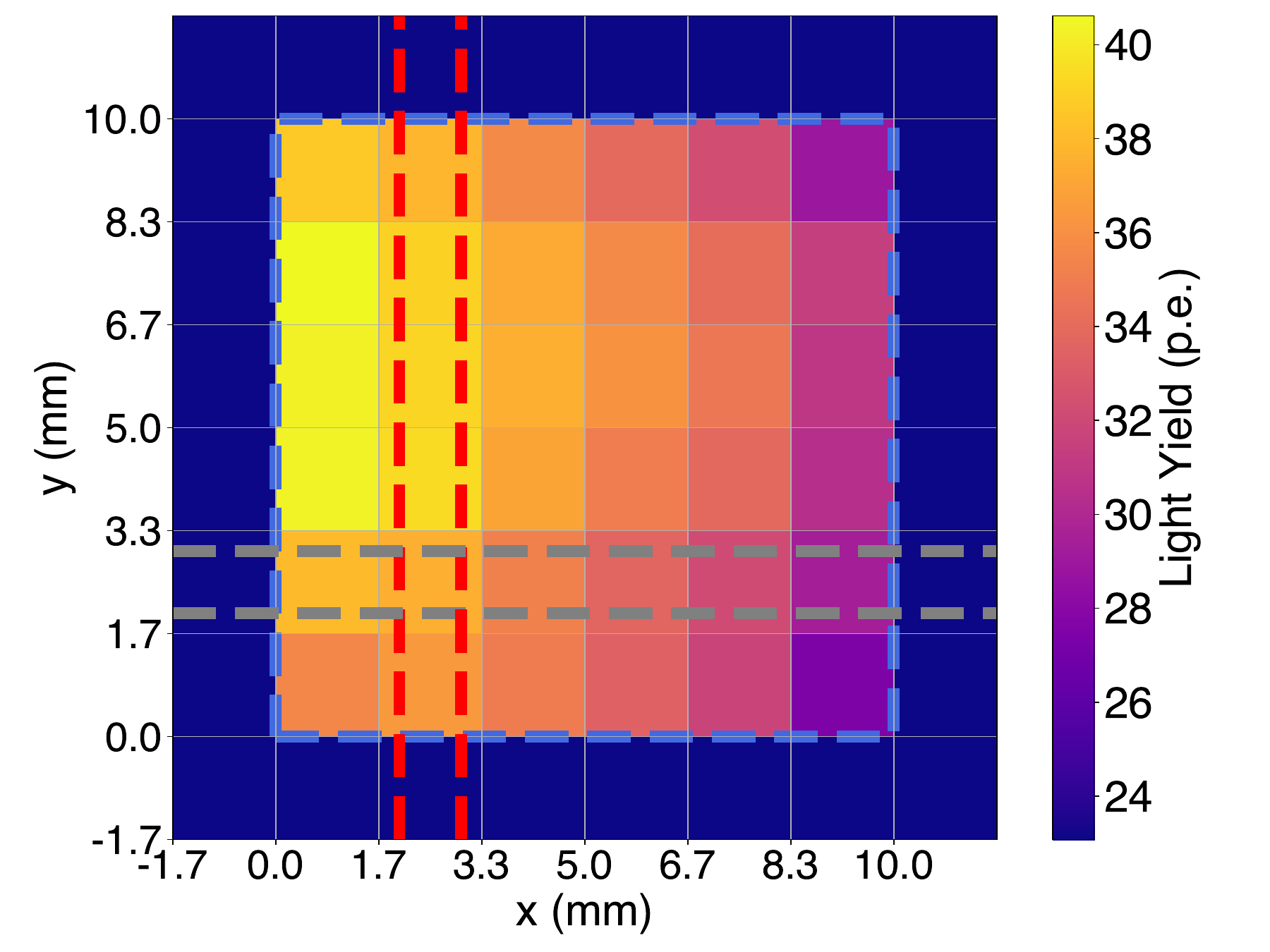}
        \caption{The readout of the WLS fibers along the Y axis.}
    \end{subfigure}
	\caption{The LY non-uniformity map within a single cube readout by either the x (top) or y (bottom) fiber channels. The boundary of the scintillator cube is shown by the blue dashed line, while the position of the readout fiber is highlighted by the red dashed lines. The position of the other fiber is shown by the grey dashed lines.} 
	\label{fig:non_uni}
\end{figure}

Figure~\ref{fig:non_uni} presents the $6\times 6$-bin LY distributions as a function of the track segment position within a single cube, for the x fibers (top) and the y fibers (bottom), averaged over all the five cubes of the center row. Each map corresponds to the the same area as a center column of the prototype of size 1 cm$^2$. The color of each bin represents the average LY measured for tracks passing through that region in the cube. The edge of the cube is highlighted with blue dashed lines, while the active fiber for each map is highlighted in red, and the orthogonal fiber in gray. The maps clearly show higher light-yield regions around the WLS fibers. As expected, the closer the ionization position is to a fiber, the greater the detected LY. Shadowing effects between the two WLS fibers are also visible: particles closer to one fiber produce less light detected by the orthogonal fiber.
Defined as the relative deviation of the regional LY from the overall average LY, the LY non-uniformity in the x channel ranges from -19\% to 12\%, while in the y channel it ranges from -16\% to 12\%. The overall root-mean-square (RMS) non-uniformity within a single cube was measured to be approximately 7\%.

\begin{figure}[h!]
	\centering 
    \begin{subfigure}{0.45\textwidth}
    \centering
	   \includegraphics[width=\linewidth]{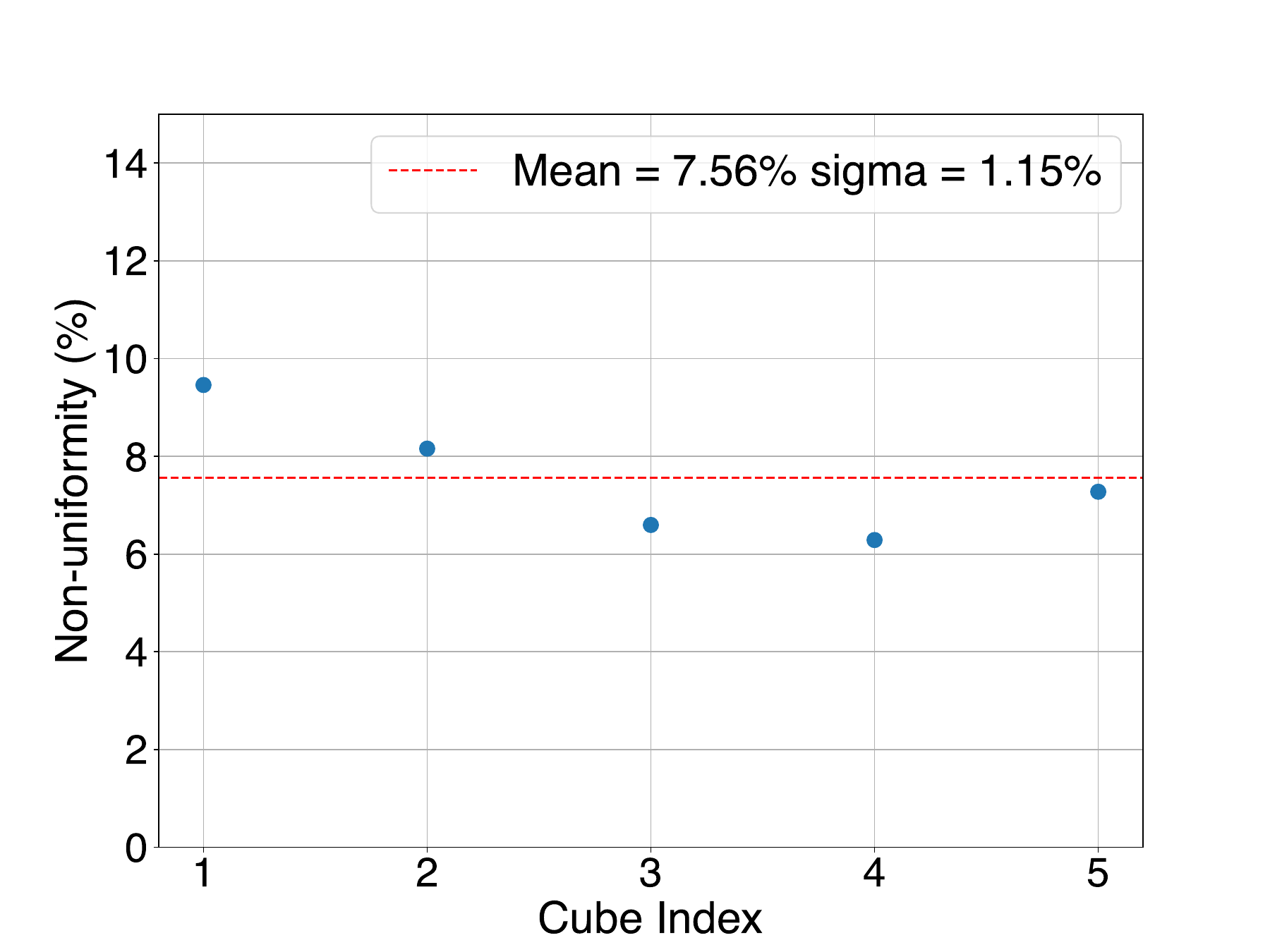}	
        \caption{x fiber channel}
    \end{subfigure}%
    \newline
    \centering
    \begin{subfigure}{0.45\textwidth}
    \centering
        \includegraphics[width=\linewidth]{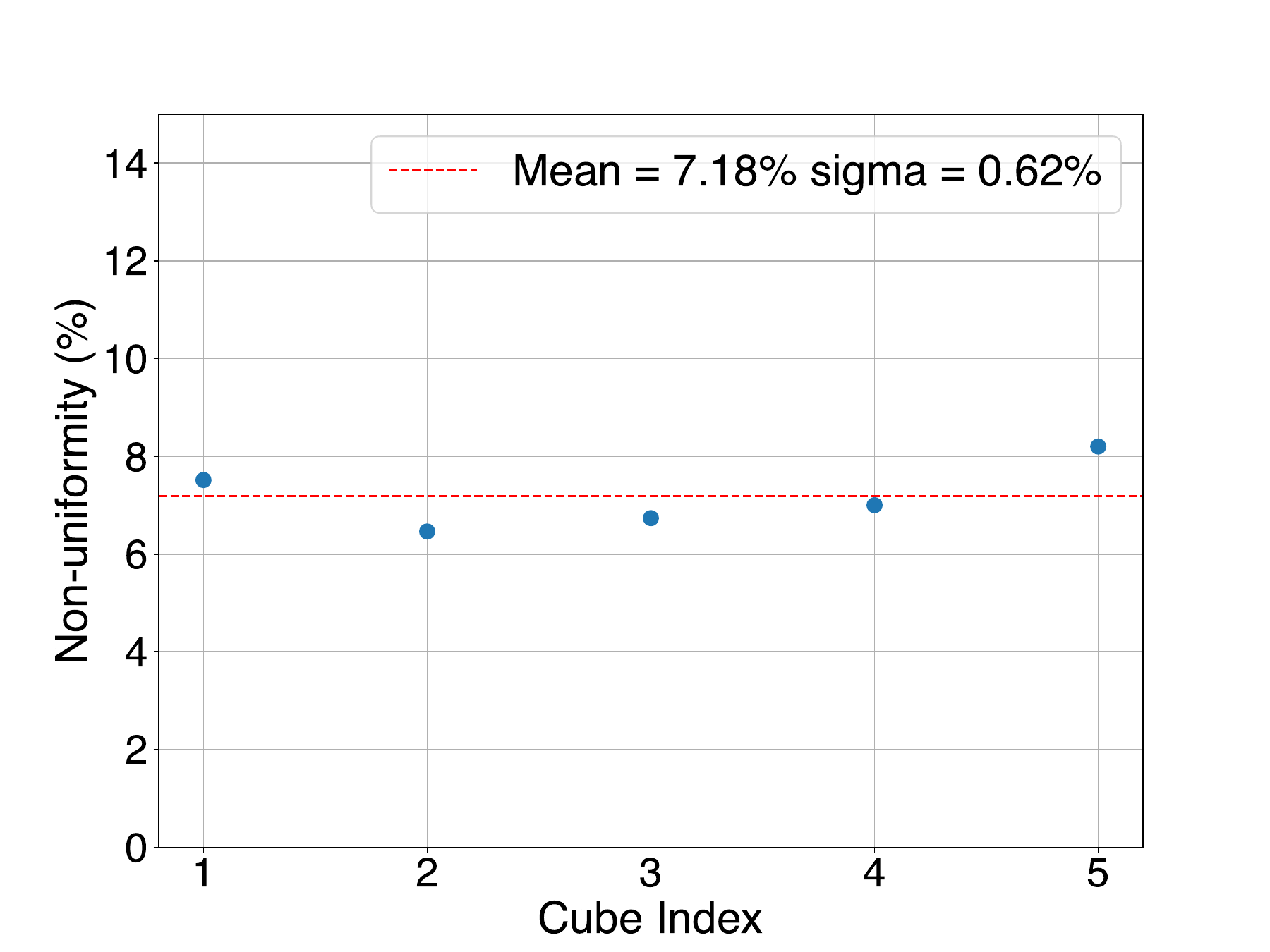}
        \caption{y fiber channel}
    \end{subfigure}
	\caption{The LY non-uniformity is derived as the RMS of the LY measured at different particle interaction positions, as shown in Fig.~\ref{fig:non_uni}. The measured non-uniformity of the five cubes in the central row covered by the hodoscoped is plotted above and the average is shown by the dashed red line  (Top: x channel. Bottom: y channel). On average the non-uniformity within a single cube is about 7\% with standard deviation of about 0.8\% between different cubes.} 
	\label{fig:non_uni_non_uni}
\end{figure}

The analysis was extended to all five cubes in the central row, which were covered by the hodoscopes. For each cube, the RMS of the LY non-uniformity was calculated from its non-uniformity map, and the results are summarized in Fig.~\ref{fig:non_uni_non_uni}. An average RMS non-uniformity of approximately 7\% was observed, with low variation ($\sim$ 0.9\%) across different cubes. This shows the good uniformity in the response of the 3D-printed scintillator cubes, as well as an excellent reproducibility.

\section{Conclusions}
The first prototype 3D-printed using the FIM technique, consisting of 125 plastic scintillator voxels, each of 1 $\times$ 1 $\times$ 1 cm$^3$ size, and read out by two orthogonal WLS fibers, was successfully tested in a charged particle beam at the Proton-Synchrotron facility at CERN. In total 50 WLS fiber channels were instrumented, to provide 3D particle tracking.
Two hodoscopes were installed at both side of the SuperCube, covering the whole cross section of the central cubes, enabling a high spatial resolution study of the LY non-uniformity within a single voxel.
A typical LY of about 28 p.e. per channel was measured, being consistent with the result reported in \cite{FIM} obtained with cosmic muons, and comparable with a sample manufactured by standard cast polymerization \cite{glued_cube}. The cube-to-cube light leakage was measured to be 4\% - 5\% on average, meaning that in total about 24\% - 30\% of the detectable scintillating light escapes through reflective walls. 
The spatial non-uniformity of the detected LY were measured from the center 5 voxels covered by the two hodoscopes. A LY non-uniformity depending on the proximity of the interacting point to the WLS fibers is measured to be about 7\% within a single cube. Less than 1\% of variation has been observed by comparing the five cubes in the center column, showing the nice stability and reproducibility of the FIM manufactured plastic scintillator voxels.
A new reflective filament is under developing and will potentially further reduce the light leakage of the prototype and increase the detected LY. Work is in progress to add the WLS fiber holes for the third readout view.

\section*{Acknowledgements}
This work was supported by the SNSF grant PCEFP2 203261, Switzerland.




\bibliographystyle{JHEP}
\bibliography{SuperCube_Beamtest}

\providecommand{\href}[2]{#2}\begingroup\raggedright\begin{thebibliography}{10}

\bibitem{PhysRev.80.474}
M.~G. Schorr and F.~L. Torney, \emph{Solid non-crystalline scintillation phosphors}, \href{http://dx.doi.org/10.1103/PhysRev.80.474}{\emph{Phys. Rev.} {\bfseries 80} (Nov, 1950) 474--474}.

\bibitem{Amaudruz:2012esa}
P.-A. Amaudruz et~al., \emph{{The T2K Fine-Grained Detectors}}, \href{http://dx.doi.org/10.1016/j.nima.2012.08.020}{\emph{Nucl. Instrum. Meth. A} {\bfseries 696} (2012) 1--31}, [\href{https://arxiv.org/abs/1204.3666}{{\ttfamily 1204.3666}}].

\bibitem{Aliaga:2013uqz}
{\scshape MINERvA} collaboration, L.~Aliaga et~al., \emph{{Design, Calibration, and Performance of the MINERvA Detector}}, \href{http://dx.doi.org/10.1016/j.nima.2013.12.053}{\emph{Nucl. Instrum. Meth. A} {\bfseries 743} (2014) 130--159}, [\href{https://arxiv.org/abs/arXiv:1305.5199}{{\ttfamily arXiv:1305.5199}}].

\bibitem{MINOS:2008hdf}
{\scshape MINOS} collaboration, D.~G. Michael et~al., \emph{The magnetized steel and scintillator calorimeters of the minos experiment}, \href{http://dx.doi.org/10.1016/j.nima.2008.08.003}{\emph{Nucl. Instrum. Meth. A} {\bfseries 596} (2008) 190--228}, [\href{https://arxiv.org/abs/0805.3170}{{\ttfamily 0805.3170}}].

\bibitem{Joram:2015ymp}
C.~Joram, U.~Uwer, B.~D. Leverington, T.~Kirn, S.~Bachmann, R.~J. Ekelhof et~al., \emph{{LHCb Scintillating Fibre Tracker Engineering Design Review Report: Fibres, Mats and Modules}},  2015.

\bibitem{TheATLASCollaboration_2008}
T.~A. Collaboration, \emph{{The ATLAS Experiment at the CERN Large Hadron Collider}}, \href{http://dx.doi.org/10.1088/1748-0221/3/08/S08003}{\emph{Journal of Instrumentation} {\bfseries 3} (aug, 2008) S08003}.

\bibitem{Betancourt2017}
C.~Betancourt, A.~Blondel, R.~Brundler, A.~Daetwyler, Y.~Favre, D.~Gascon et~al., \emph{{Application of large area SiPMs for the readout of a plastic scintillator based timing detector}}, \href{http://dx.doi.org/10.1088/1748-0221/12/11/P11023}{\emph{Journal of Instrumentation} {\bfseries 12} (2017) P11023}, [\href{https://arxiv.org/abs/arXiv:1709.08972}{{\ttfamily arXiv:1709.08972}}].

\bibitem{Korzenev:2021mny}
A.~Korzenev et~al., \emph{{A 4\ensuremath{\pi} time-of-flight detector for the ND280/T2K upgrade}}, \href{http://dx.doi.org/10.1088/1748-0221/17/01/P01016}{\emph{Journal of Instrumentation} {\bfseries 17} (2022) P01016}, [\href{https://arxiv.org/abs/2109.03078}{{\ttfamily 2109.03078}}].

\bibitem{Allan:2013ofa}
D.~Allan et~al., \emph{{The Electromagnetic Calorimeter for the T2K Near Detector ND280}}, \href{http://dx.doi.org/10.1088/1748-0221/8/10/P10019}{\emph{Journal of Instrumentation} {\bfseries 8} (2013) P10019}, [\href{https://arxiv.org/abs/1308.3445}{{\ttfamily 1308.3445}}].

\bibitem{calice}
V.~Andreev et~al., \emph{A high-granularity plastic scintillator tile hadronic calorimeter with apd readout for a linear collider detector}, \href{http://dx.doi.org/10.1016/j.nima.2006.04.044}{\emph{Nucl. Instrum. Meth. A} {\bfseries 564} (2006) 144--154}.

\bibitem{SoLid:2017ema}
{\scshape SoLid} collaboration, Y.~Abreu et~al., \emph{A novel segmented-scintillator antineutrino detector}, \href{http://dx.doi.org/10.1088/1748-0221/12/04/P04024}{\emph{Journal of Instrumentation} {\bfseries 12} (2017) P04024}, [\href{https://arxiv.org/abs/1703.01683}{{\ttfamily 1703.01683}}].

\bibitem{Sgalaberna:2017khy}
A.~Blondel et~al., \emph{A fully-active fine-grained detector with three readout views}, \href{http://dx.doi.org/10.1088/1748-0221/13/02/p02006}{\emph{Journal of Instrumentation} {\bfseries 13} (feb, 2018) P02006--P02006}.

\bibitem{Mineev:2018ekk}
O.~Mineev et~al., \emph{{Beam test results of 3D fine-grained scintillator detector prototype for a T2K ND280 neutrino active target}}, \href{http://dx.doi.org/10.1016/j.nima.2019.01.080}{\emph{Nucl. Instrum. Meth. A} {\bfseries 923} (Apr, 2019) 134–138}.

\bibitem{sfgd-testbeam-cern}
A.~Blondel et~al., \emph{{The SuperFGD Prototype Charged Particle Beam Tests}}, \href{http://dx.doi.org/10.1088/1748-0221/15/12/P12003}{\emph{Journal of Instrumentation} {\bfseries 15} (2020) P12003}, [\href{https://arxiv.org/abs/arXiv:2008.08861}{{\ttfamily arXiv:2008.08861}}].

\bibitem{ND280upgrade-tdr}
{\scshape T2K} collaboration, K.~Abe et~al., ``{T2K ND280 Upgrade - Technical Design Report}.'' https://arxiv.org/abs/1901.03750, 2020.

\bibitem{nd280upgrade-press-release}
``T2k press release.'' \url{https://www.kek.jp/en/press-en/202401171405/}.

\bibitem{Fedotov:2021ylh}
S.~Fedotov, A.~Dergacheva, A.~Filik, M.~Khabibullin, A.~Khotjantsev, Y.~Kudenko et~al., \emph{{Scintillator cubes for 3D neutrino detector SuperFGD}}, \href{http://dx.doi.org/10.1088/1742-6596/2374/1/012106}{\emph{J. Phys. Conf. Ser.} {\bfseries 2374} (2022) 012106}, [\href{https://arxiv.org/abs/2111.07305}{{\ttfamily 2111.07305}}].

\bibitem{Alonso-Monsalve:2022zlm}
S.~Alonso-Monsalve, D.~Sgalaberna, X.~Zhao, C.~McGrew and A.~Rubbia, \emph{{Artificial intelligence for improved fitting of trajectories of elementary particles in dense materials immersed in a magnetic field}}, \href{http://dx.doi.org/10.1038/s42005-023-01239-4}{\emph{Commun. Phys.} {\bfseries 6} (2023) 119}, [\href{https://arxiv.org/abs/2211.04890}{{\ttfamily 2211.04890}}].

\bibitem{fdm}
``Fdm technology, about fused deposition modeling (copyright by stratasys).'' \url{http://www.stratasys.com/3d-printers/technologies/fdm-technology}.

\bibitem{Berns:2020ehg}
S.~Berns, A.~Boyarintsev, S.~Hugon, U.~Kose, D.~Sgalaberna, A.~De~Roeck et~al., \emph{A novel polystyrene-based scintillator production process involving additive manufacturing}, \href{http://dx.doi.org/10.1088/1748-0221/15/10/P10019}{\emph{Journal of Instrumentation} {\bfseries 15} (2020) 10}, [\href{https://arxiv.org/abs/2011.09859}{{\ttfamily 2011.09859}}].

\bibitem{3DET:2022dkw}
{\scshape 3DET} collaboration, S.~Berns et~al., \emph{Additive manufacturing of fine-granularity optically-isolated plastic scintillator elements}, \href{http://dx.doi.org/10.1088/1748-0221/17/10/P10045}{\emph{Journal of Instrumentation} {\bfseries 17} (2022) P10045}, [\href{https://arxiv.org/abs/2202.10961}{{\ttfamily 2202.10961}}].

\bibitem{FIM}
T.~Weber, A.~Boyarintsev, U.~Kose, B.~Li, D.~Sgalaberna, T.~Sibilieva et~al., \emph{Additive manufacturing of a 3d-segmented plastic scintillator detector for tracking and calorimetry of elementary particles},  2024.

\bibitem{glued_cube}
A.~Boyarintsev, A.~D. Roeck, S.~Dolan, A.~Gendotti, B.~Grynyov, U.~Kose et~al., \emph{Demonstrating a single-block 3d-segmented plastic-scintillator detector}, \href{http://dx.doi.org/10.1088/1748-0221/16/12/P12010}{\emph{Journal of Instrumentation} {\bfseries 16} (dec, 2021) P12010}.

\bibitem{hamamatsu:mppc}
Hamamatsu, \emph{MPPC (multi-pixel photon counter) S13360 series}, 10, 2022.

\bibitem{kuraray-catalogue-wls}
``{Kuraray catalogue - plastic scintillating fibers}.'' \url{http://kuraraypsf.jp/psf/ws.html}.

\bibitem{caen-FERS-board}
``{CAEN FERS DT5202}.'' \url{https://www.caen.it/products/dt5202/}.

\bibitem{ENUBET}
F.~Iacob, F.~Acerbi, I.~Angelis, M.~Bonesini, A.~Branca, C.~Brizzolari et~al., \emph{{ENUBET: a monitored neutrino beam for the precision era of neutrino physics}}, \href{http://dx.doi.org/10.1088/1742-6596/2156/1/012234}{\emph{Journal of Physics: Conference Series} {\bfseries 2156} (dec, 2021) 012234}.

\end{thebibliography}\endgroup






\end{document}